\documentstyle[sprocl,epsfig]{article}
\def\PLB{{\em Phys. Lett.}  B}

\def\PRD{{\em Phys. Rev.} D}
\def\EPJ{{\em Eur. Phys. J.} C }
\def\be{\begin{equation}}
\def\ee{\end{equation}}
\def\bea{\begin{eqnarray}}
\def\eea{\end{eqnarray}}

\begin{document}
\title{TOTAL CROSS-SECTIONS : CROSS-TALK BETWEEN  HERA, LHC AND LC \\}

\author{G. PANCHERI}
\address{INFN, Frascati National Laboratories, I00044Frascati, Italy}
\author{A. DE ROECK}
\address{CERN, Geneva, Switzerland}
\author{R.M. GODBOLE}
\address{ Centre for High Energy Physics, Indian Institute of Science,
Bangalore,  560 012, India}
\author{A. GRAU}
\address{Departamento de F\'\i sica Te\'orica y del Cosmos, 
Universidad de Granada, Spain}
\author{Y.N. SRIVASTAVA}
\address{Dipartimento di Fisica \& INFN, University of Perugia,
 Perugia, Italy}

%%%%%%%%%%%%%%%%%%%%%%%%%%%%%%%%%%%%%%%%%%%%%%%%%%%%%%%%%%%%%%

% You may repeat \author \address as often as necessary      %

%%%%%%%%%%%%%%%%%%%%%%%%%%%%%%%%%%%%%%%%%%%%%%%%%%%%%%%%%%%%%%
%%%%%%%%%%%%%%%%%%%%%%%%%%%%%%%%%%%%%%%%%%%%%%%%%%%%%%%%%%%%%
%  Adding the additinal page to upload it to the web
%%%%%%%%%%%%%%%%%%%%%%%%%%%%%%%%%%%%%%%%%%%%%%%%%%%%%%%%%%%%
\begin{flushright}
{hep-ph/0412189}\\
\end{flushright}
\vspace{.5cm}
\begin{center}
{\Large \bf Total cross-sections: \\[2mm] Cross-talk between HERA, LHC and LC}
\footnote{Presented by
G. Pancheri at the International Linear Collider Workshop, April 2004,
Paris.} \\[0.3cm]
\end{center}
\begin{center}
{\large G. Pancheri$^{1}$, A. De Roeck$^{2}$ R.M. Godbole$^{3}$
A. Grau$^{4}$ and Y. N. Srivastava $^{5}$}
\\[0.35 cm]
$^1$INFN, Frascati National Laboratories, I00044, Frascati, Italy 
\\[0.20cm]
$^2$CERN, EP-Division, Geneva, Switzerland.\\[0.20cm]
$^3$Centre for High Energy Physics, Indian Institute of Science, 
Bangalore, 560012, India.\\[0.20cm]
$^4$Departamento de F\'\i sica Te\'orica y del Cosmos,
Universidad de Granada, Spain.\\[0.20cm]
$^5$Dipartimento di Fisica \& INFN, University of Perugia,
 Perugia, Italy.
\end{center}

\vspace{.4cm}

\begin{abstract}
{\noindent\normalsize
We discuss the need to compare
total cross-section measurements at LHC and HERA with each other and with
available models in order to obtain  a more precise prediction
of the total hadronic cross-section at the future Linear Collider, thus
leading to a better estimate of the hadronic background.
}
\end{abstract}
\newpage
\maketitle
\abstracts{We discuss the need to compare 
 total cross-section measurements at LHC and HERA with each other and with
available models in order to obtain  a more precise prediction 
of the total hadronic cross-section at the future Linear Collider, thus 
leading to a better estimate of the hadronic background.}
%**********************************************************************

\section{Introduction}

%***********************************************************************

In order to make realistic predictions for
 $e^+e^- \rightarrow hadrons $ 
at the Linear Collider (LC), one needs to understand the role played by 
QCD in the energy dependence of total cross-sections. This will allow to
reduce the large errors coming from the present large uncertainties in
the $\gamma \gamma \rightarrow hadrons$ and which affect present
predictions at the LC. 

\section{Hadronic Total Cross-sections}
\begin{figure}[ht] 
 \centerline{
      \epsfig{file=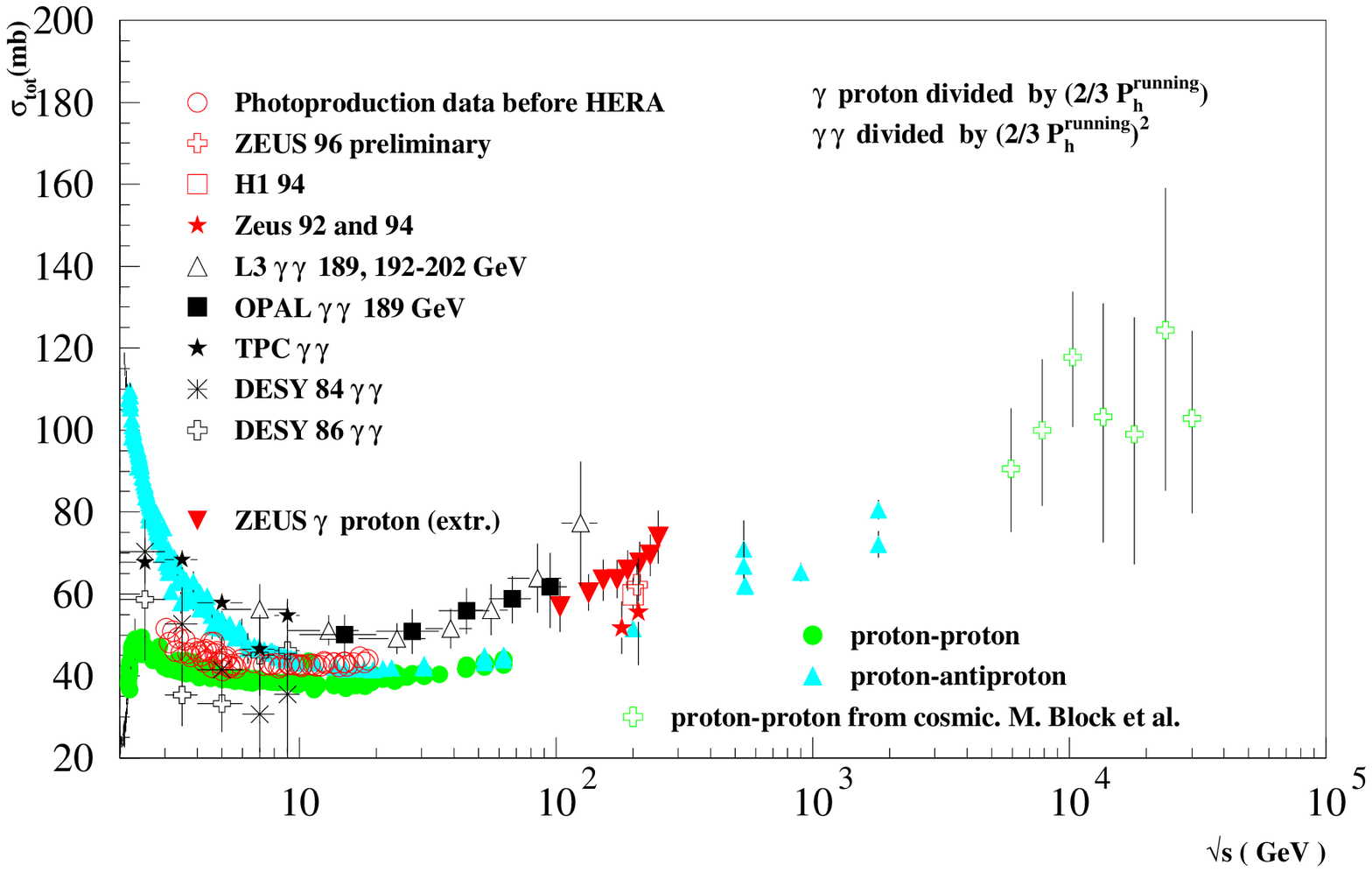,width=7cm,height=6cm}
      \epsfig{file=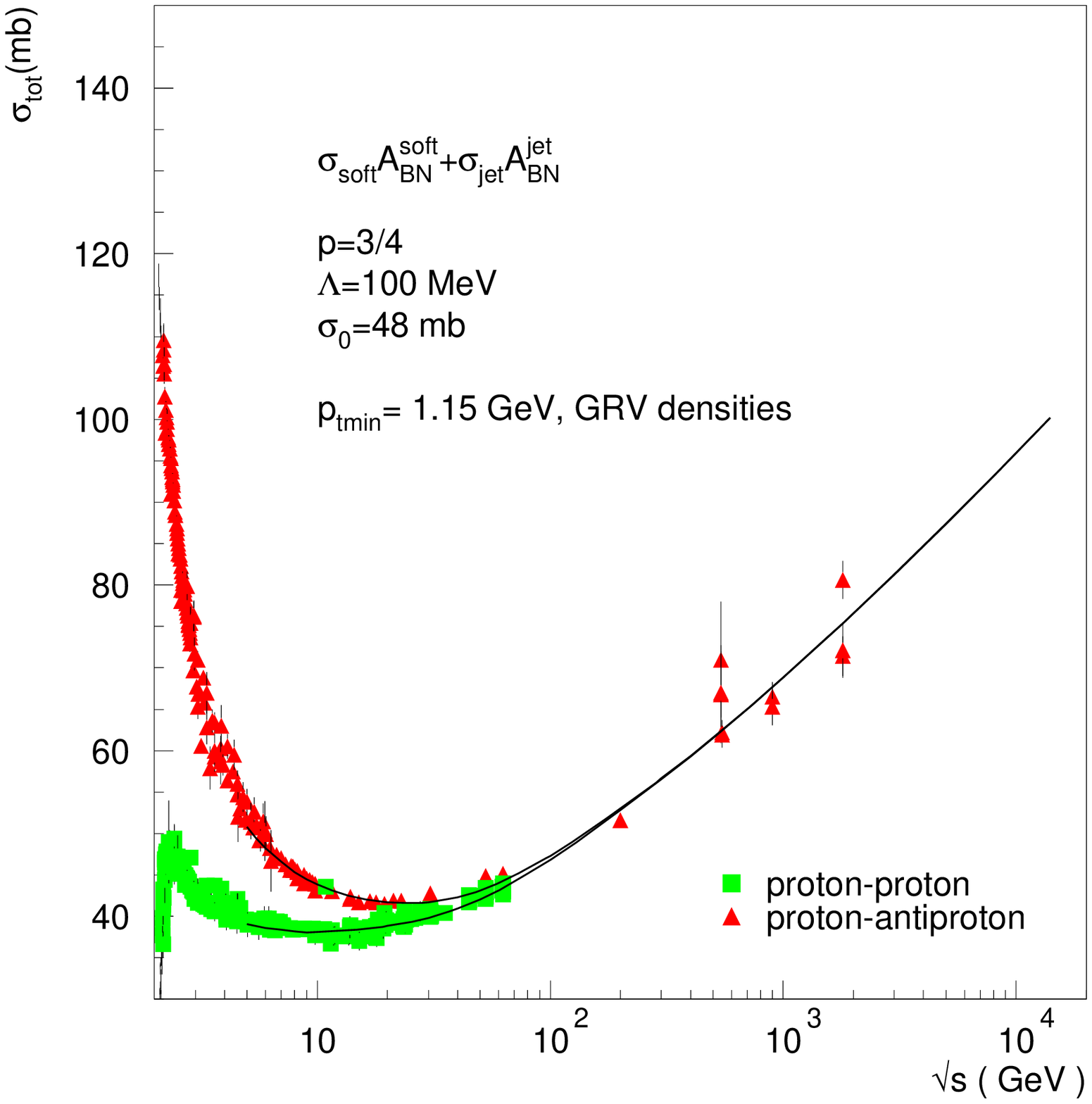,width=5.7cm,height=6cm}}
\vspace{-.4cm}
\caption{At left, proton and photon total cross-sections; those 
for photons  scaled  by a quark counting factor ${{2}\over{3}}$ 
and a VMD factor. At right the
total $p p$ and $p\bar{p}$ cross-section compared to a QCD model with
minijets  in the eikonal formulation and soft
gluons to soften the rise.} 
\label{fig:allfig1}
\end{figure}

It must be noticed that presently all total
cross-sections exhibit the same general features, namely an initial
decrease followed by a more or less gentle rise\cite{DL}, as shown in 
Fig.(\ref{fig:allfig1}). We  however lack a
complete theoretical model and while we might be able to do a good
parametrization of the proton-proton data, we do not know how to gauge the
inherent theoretical uncertainties. This results in further uncertainties
in the photoproduction cross-section and in an even larger uncertainty
in predictions at LC. Thus we should study in detail the cross-talk
between future measurements of total proton-proton cross-section at the LHC 
and a similar HERA measurement, to develop and fully understand what to expect
at LC in terms of the hadronic background.

\begin{figure}[ht]
\centerline{
\epsfig{file=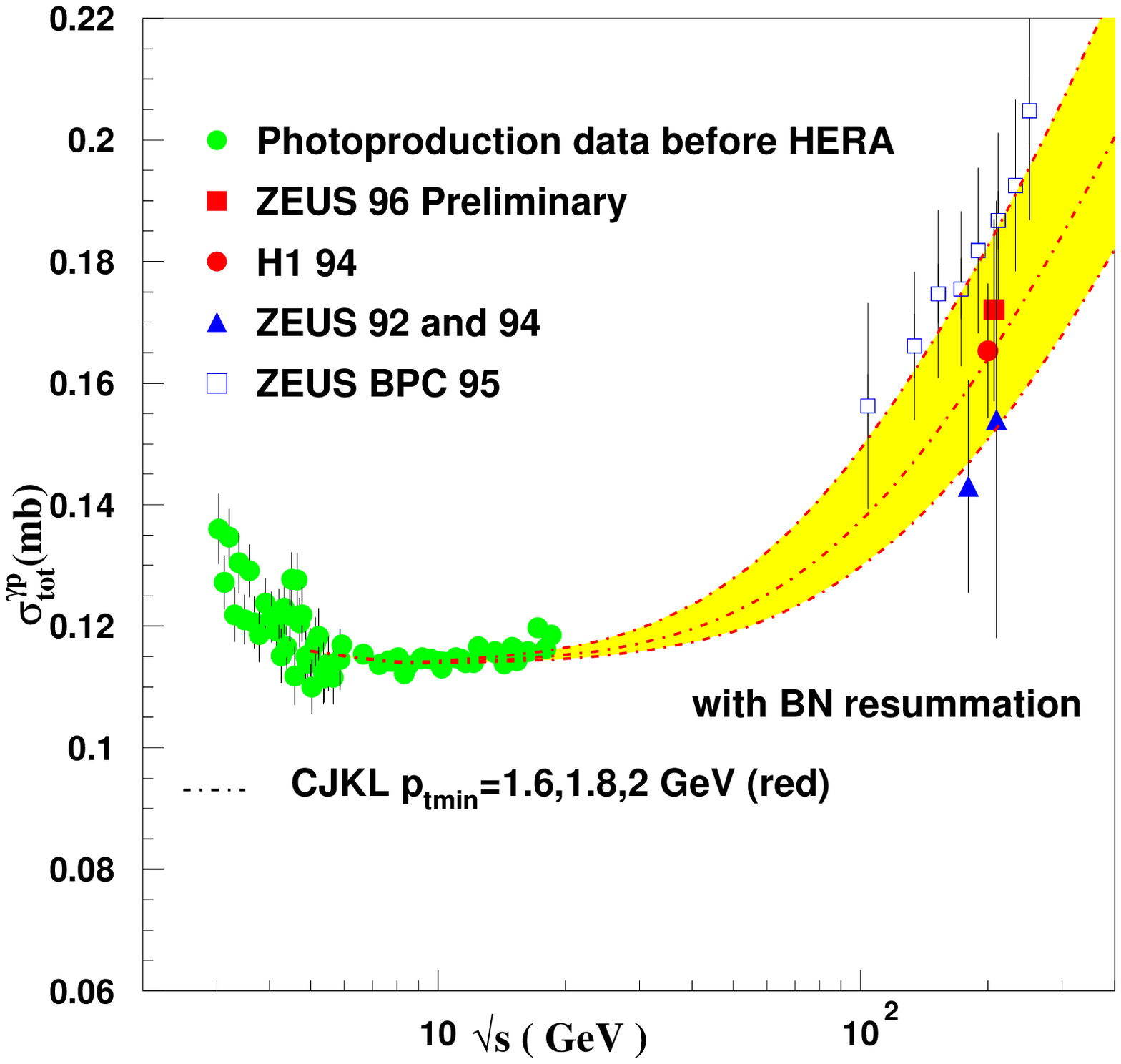,width=6.0cm,height=5cm}
\epsfig{file=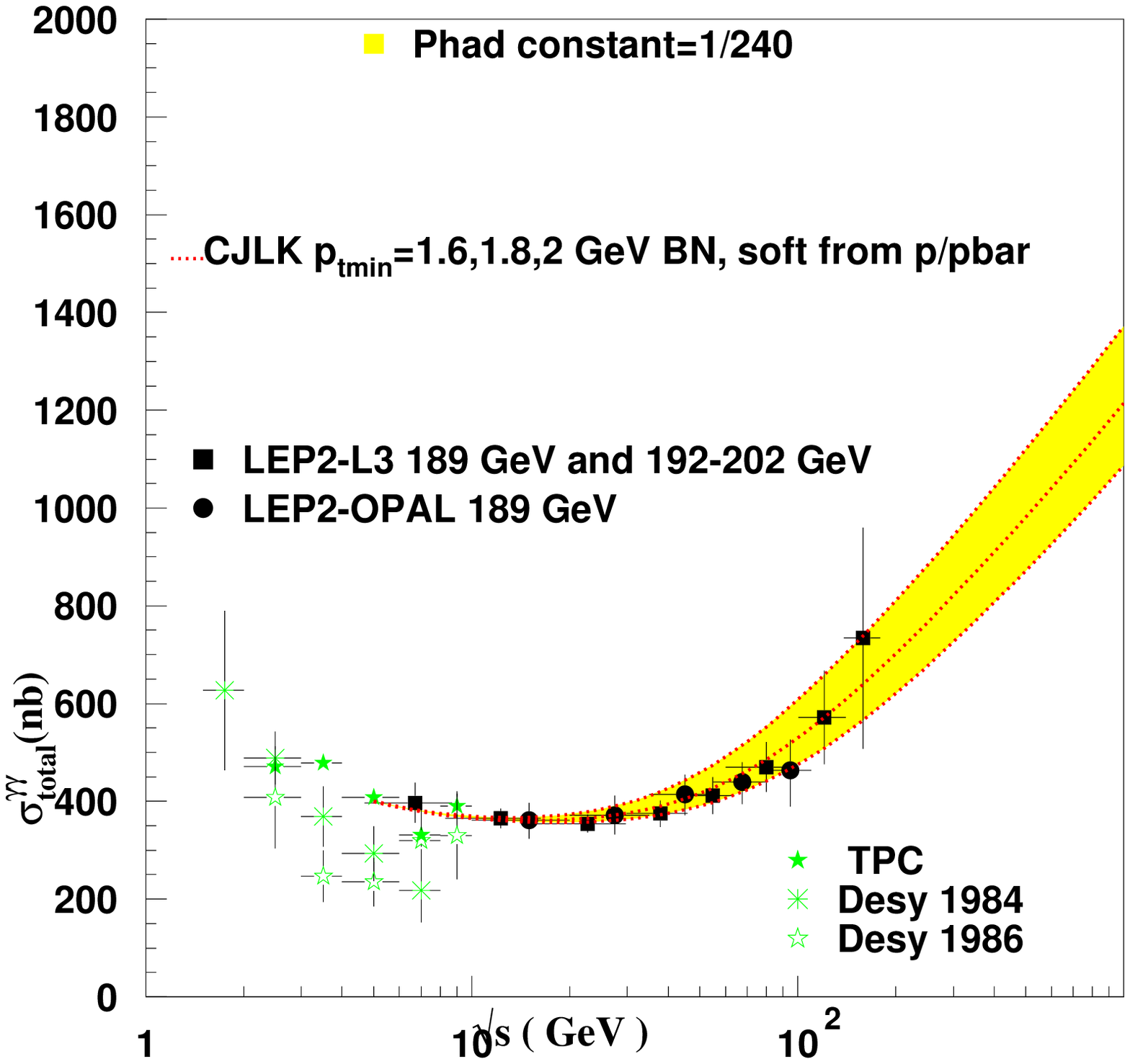,width=6.5cm,height=6cm}}
\caption{At left we show  photoproduction data compared with the soft
     gluon improved EMM, at right the same for $\gamma \gamma$, using the
     same photon densities and set of $p_{tmin}$ values.}
     \label{fig:gggpfig2}
     \end{figure}

The curve shown in Fig.(\ref{fig:allfig1}) is obtained by embedding the QCD 
mini-jet cross-section in an eikonal formulation
 \cite{ourPRD}  
  \begin{equation}
\sigma_{tot}=2\int d^2{\vec b}[1-e^{i\chi(b,s)}]
\label{stot}
\end{equation}
where we put  $\chi_R\approx 0$, and 
\begin{equation}
2\chi_I=n(b,s)=A_{BN}^{soft}\sigma_{soft}+A_{BN}^{jet}\sigma_{jet}
\end{equation}
In this Eikonal Minijet Model (EMM), the rise is driven by the 
jet cross-section, which is  calculated using current parton densities 
down to a minimum jet transverse momentum, $p_{tmin} \approx 1 \div 2\
GeV$.
The impact parameter distribution is the (normalized to unity)
Fourier transform of the initial transverse momentum of the colliding
partons, in leading order the valence quarks, obtained through the 
Bloch Nordsieck soft gluon summation\cite{ourPRD}.  
This model can also be applied to the photo-production and $\gamma \gamma$
data. The results, for a set of $p_{tmin}$ values, are shown in 
Fig. (\ref{fig:gggpfig2}), with CJKL \cite{cjkl} parton densities.
\begin{figure}[ht]
   \centerline{
   \epsfig{file=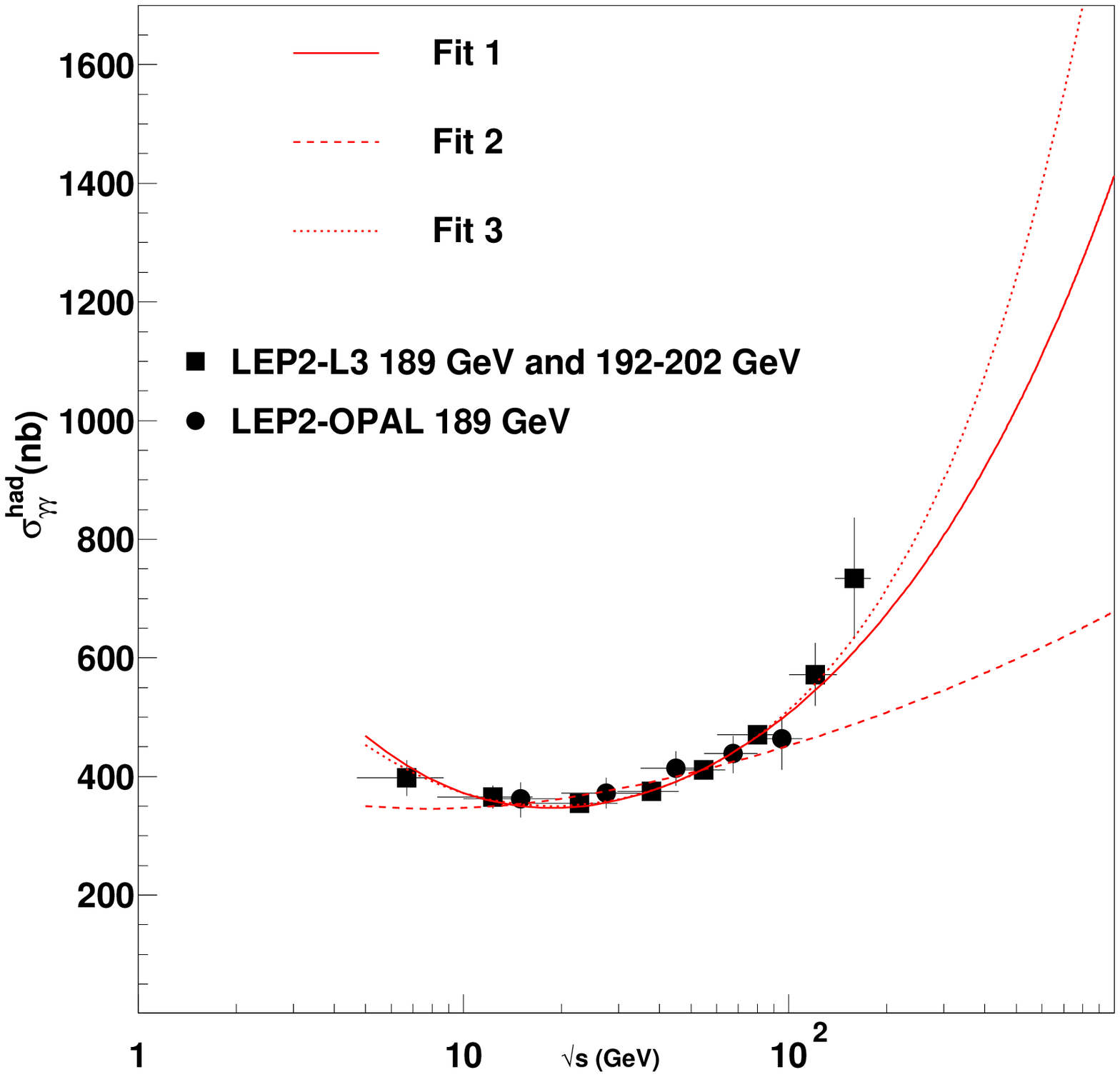,width=6.5cm,height=6cm}
   \epsfig{file=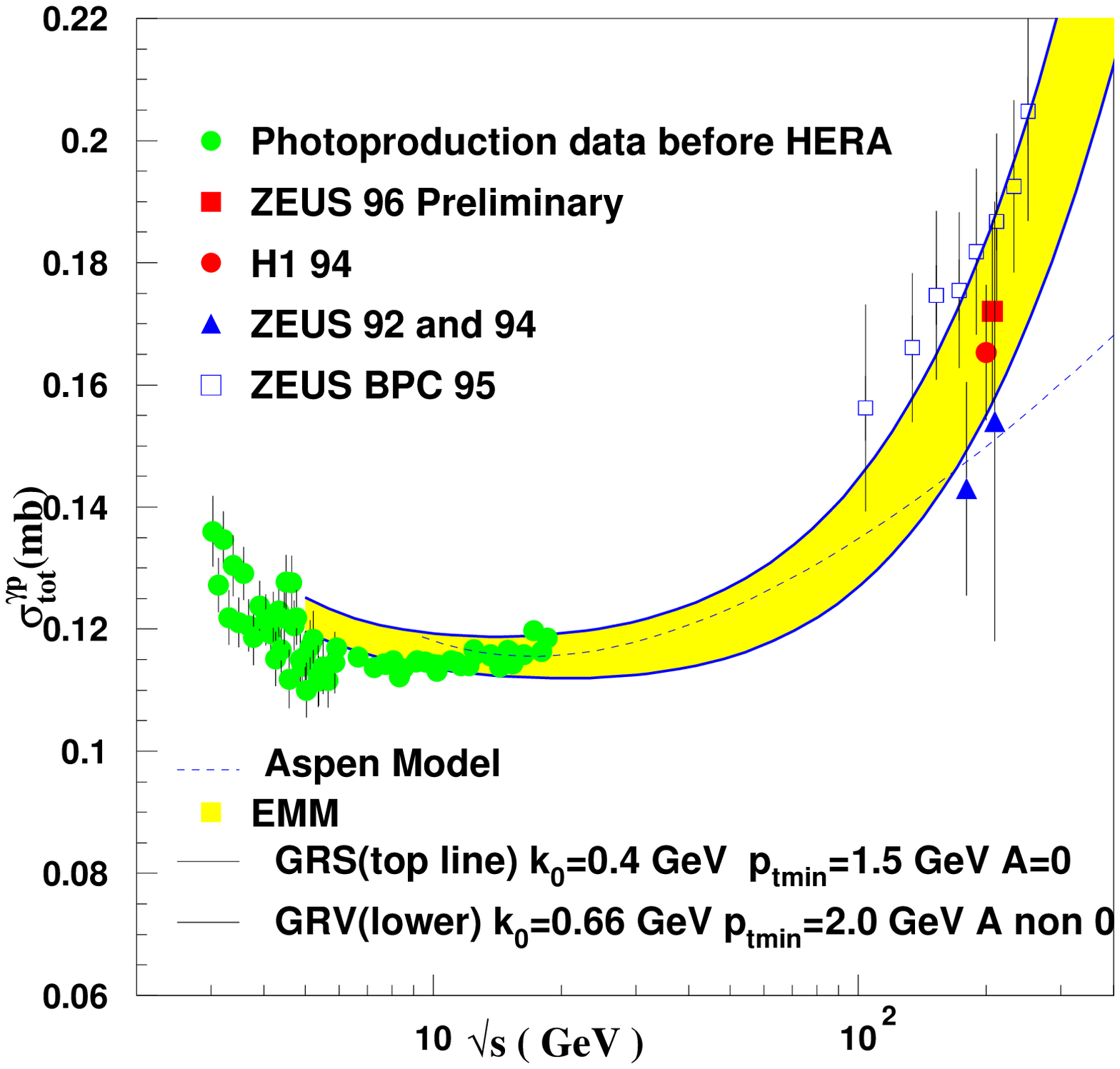,width=7.0cm,height=6cm}}
\caption{At left we show  how  different power laws can be used to 
parametrize $\gamma \gamma \rightarrow hadrons$ with fit 2 
corresponding to the same slope parameters as in proton-proton, 
fit 1 to  a rising power characterized by
 $\epsilon=0.250$ and fit 3 includes two rising powers, one as 
in proton-proton, and a second one with $\epsilon'=0.418$. At right we 
show photoproduction data fitted through the Aspen model (with same gentle
rise as proton-proton) or through the EMM with Form Factors for the 
impact-parameter distribution.  }
     \label{fig:gggpfig3}
     \end{figure}

The EMM model with soft gluons has the virtue of reproducing the gentle rise
with energy which characterizes the proton-proton data and is also
compatible with the photon data. This is not true for other models,
as we show in Fig.(\ref{fig:gggpfig3}), where a  fit \cite{ourwithalbert} 
of the $\gamma \gamma $ data with a Regge-Pomeron type parametrization, 
indicates a harder rise with energy than what is present in the 
proton-proton data. We also show the photoproduction data and compare it 
with the (i) EMM model without soft gluons \cite{pancheri} and with 
the (ii) Aspen model~\cite{aspen}.

\section{Conclusion}
Present Tevatron data for total cross-section lead to  uncertainties
when extrapolated to the LHC energies. Such uncertainties 
are also found in predictions for LC hadronic cross-sections which suffer 
from uncertainties in the $\gamma\gamma$ cross-sections. We have emphasized
that to obtain realistic estimates for the hadronic background at LC, 
we need combined model predictions for LHC as well as the HERA data.
An example has been provided through our soft gluon formula for the
purely hadronic as well as the $\gamma\gamma$ data.
\section*{Acknowledgments}
This work was supported in part by EU Contract HPRN-CT2002-00311.
RG wishes to acknowledge the partial support of the Department of
Science and Technology, India, under project number SP/S2/K-01/2000-II.
AG acknowledges support from MCYT under project number FPA2003-09298-c02-01.


\section*{References}
\begin{thebibliography}{99}
\bibitem{DL} {A. Donnachie  and P.V. Landshoff,
\PLB  \rm {\bf 296} (1992)  227.}\\
\bibitem{ourPRD} {R.M. Godbole, A. Grau, G. Pancheri and Y.N. Srivastava, 
hep-ph/0408355, and references therein.}\\
\bibitem{cjkl} {F. Cornet, P. Jankowski, M. Krawczyk and A. Lorca, 
\PRD \rm {\bf 68} (2003) 014010, hep-ph/0212160.}\\
\bibitem{ourwithalbert}{R.M. Godbole, A. de Roeck, A. Grau and G. Pancheri, 
{\em JHEP} 0306:061,2003, hep-ph/0305071.}\\

\bibitem{pancheri} {R. M. Godbole,  G. Pancheri,
 \EPJ \rm {\bf 19} (2001) 129, hep-ph/0010104.}\\
\bibitem{aspen} {M.M. Block, E.M. Gregores, F. Halzen and  G. Pancheri,
 \PRD \rm {\bf 58} (1998) 17503; M. Block, E.M. Gregores, F. Halzen
and G. Pancheri, \PRD \rm {\bf 60} (1999) 54024.}



\end{thebibliography}
\end{document}